\documentclass{article}
\usepackage{epsfig}
\usepackage{amsmath}
\usepackage{amssymb}
\usepackage{amsfonts}
\usepackage{enumerate}
\usepackage{cite}
\newcommand{\Hrule}{\vskip-10mm\begin{flushleft}\rule{\linewidth}{0.1mm}\end{flu
shleft}\vskip-1.5mm}
\title{Issues pertaining to  D'yakonov-Perel' spin relaxation in quantum wire 
channels}
\author{Sandipan Pramanik$^{\dag}$, Supriyo 
Bandyopadhyay$^{\dag}$$\thanks{Corresponding author. E-mail: sbandy@vcu.edu}$, 
Marc Cahay$^{\dag \dag}$ \\
\\$^{\dag}$ Department of Electrical Engineering\\
Virginia Commonwealth University, Richmond, Virginia 23284\\
$^{\dag \dag}$Department of Electrical and Computer Engineering and Computer 
Science\\
University of Cincinnati, Cincinnati, Ohio 45221}
\begin{document}
\maketitle

\begin{abstract}

We elucidate the origin and nature of the D'yakonov-Perel' spin relaxation  in a 
quantum wire structure, showing (analytically) that there are three necessary 
conditions for it to exist: (i) transport must be multi-channeled, (ii) there 
must be a Rashba spin orbit interaction in the wire, and (iii) there must also 
be  a Dresselhaus spin orbit interaction. Therefore, the only effective way to 
completely eliminate the D'yakonov-Perel' relaxation in compound semiconductor 
channels with structural and bulk inversion asymmetry is to ensure strictly 
single channeled transport. In view of that, recent proposals in the literature 
that advocate using multi-channeled quantum wires for spin transistors appear 
ill-advised.
\end{abstract}
\bigskip

\section{Introduction}
Coherent spin transport in semiconductor quantum wires is the basis for  
interesting spintronic devices such as the Spin Field Effect Transistor 
(SPINFET) \cite{datta}. In this device (and its closely related cousins) a quasi 
one-dimensional quantum ``wire'' (as opposed to a quasi two-dimensional quantum 
``well'') is preferred as the channel for several reasons. First, one 
dimensional confinement of carriers ameliorates the harmful effects of ensemble 
averaging (at a finite temperature), thereby producing a strong conductance 
modulation \cite{datta}. This is a pre-requisite for any good ``transistor'' 
where the conductance of the ``on'' and ``off'' states must differ by several 
orders of magnitude. Second,  
one-dimensional confinement leads to a severe suppression of spin relaxation 
\cite{pramanik_prb}, \cite{pramanik_apl}. As a result, the transistor channel 
can be made long, 
which not only relaxes the demands on fabrication, but also reduces the 
threshold voltage for switching the device (the threshold voltage of a SPINFET 
is inversely proportional to the channel length). This, in turn, reduces the 
dynamic power dissipation. Of course, increasing the gate length also increases 
the transit time through the channel and the switching delay, but the power 
dissipation is proportional to the {\it square} of the threshold voltage and 
hence inversely proportional to the square of the gate length, while the transit 
time is linearly proportional to the gate length. As a result, the important 
figure of merit -- the power delay product -- scales inversely with the gate 
length. A reduced power delay product may be ultimately the most significant 
advantage that spintronics 
has over conventional electronics.

This paper is organized as follows. In the next section, we discuss the 
D'yakonov-Perel' spin relaxation in a quantum wire structure and derive 
analytical expressions for the spatial evolution of the {\it average} spin of an 
electron ensemble as a consequence of D'yakonov-Perel' relaxation. The derived 
expressions are perfectly general and are valid in the presence of arbitrary 
driving electric fields, momentum randomizing collisions and inter-subband 
scattering. Based on these expressions, we derive the necessary and sufficient 
conditions for the D'yakonov-Perel' spin relaxation to exist in a quantum wire. 
Finally, we conclude by stressing the importance of ensuring single channeled 
transport in spintronic devices in order to eliminate the D'yakonov-Perel' 
relaxation. 

\section{D'yakonov-Perel' relaxation}

The D'yakonov-Perel' spin relaxation is caused by {\it momentum-dependent} 
spin-orbit interactions that originate from bulk inversion asymmetry (giving 
rise to a Dresselhaus interaction) and structural inversion asymmetry (giving 
rise to a Rashba interaction). In this section, we will analytically derive the 
temporal and spatial evolution of the average spin of an ensemble of electrons 
in a quantum wire in the presence of these spin orbit interactions. This will 
elucidate the origin of the D'yakonov-Perel' relaxation in a quasi 
one-dimensional structure, and identify pathways to eliminate it.

\begin{figure}
\centering
\includegraphics[width=2.5in]{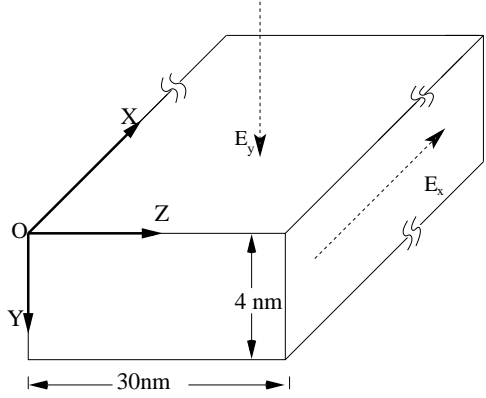}
\caption{Geometry of the quantum wire. Here $E_x$ is the longitudinal electric 
field that induces current flow. The transverse electric field $E_y$ induces 
Rashba spin-orbit coupling. }
\label{structure}
\end{figure}

Consider the quantum wire structure shown in Figure \ref{structure}. A 
transverse electric field $E_y \hat y$ is applied perpendicular to the wire axis 
($\hat x$) to induce a structural inversion asymmetry that causes a Rashba spin 
orbit interaction \cite{rashba}. This structure mimics the SPINFET \cite{datta}. 
Since materials that have strong Rashba coupling (preferred for SPINFETs) 
usually also have bulk inversion asymmetry, we will assume that there is also a 
Dresselhaus interaction \cite{dresselhaus}.

Spin evolution in the presence of spin-orbit interaction is treated by the 
standard spin density matrix \cite{saikin}
\begin{equation}
\label{spin_density}
\rho_\sigma(t)=\left[\begin{matrix} \rho\uparrow\uparrow(t)& 
\rho\uparrow\downarrow(t)\\ \rho\downarrow\uparrow(t)& 
\rho\downarrow\downarrow(t)\end{matrix}\right]
\end{equation}
which is related to the spin polarization component as 
$S_n(t)=Tr[\sigma_n\rho_\sigma(t)]$ where $n$ = $x$, $y$, $z$ and $\sigma_n$-s 
are Pauli spin matrices. This spin density matrix evolves under the influence of 
momentum dependent spin orbit coupling Hamiltonian $H_{SO}(\vec k)$ as
\begin{equation}
\label{spin_evolution}
\rho_\sigma(t+\delta t)=exp\left[-\frac{iH_{SO}(k)\delta 
t}{\hbar}\right]\rho_\sigma(t)\ exp\left[\frac{iH_{SO}(k)\delta t}{\hbar}\right]
\end{equation}
The spin-orbit coupling Hamiltonian has two main components: one due to 
Dresselhaus interaction
\begin{equation}
\label{dresselhaus1}
H_D(k)=\gamma\vec\sigma\cdot\vec\kappa
\end{equation}
and the other  due to Rashba interaction, whose strength depends on the 
transverse electric field $E_y$ and is given by
\begin{equation}
\label{rashba1}
H_R(k)=\eta\left[\vec\sigma\times\vec k\right]\cdot\hat y
\end{equation}
The constants $\gamma$ and $\eta$ depend on the material and, in case of $\eta$, 
also on the external electric field $E_y$.

In equation (\ref{dresselhaus1}), $\vec\kappa$ is given by $\vec 
\kappa=\kappa_x\hat x+\kappa_y\hat y+\kappa_z\hat z$ \cite{das} where 
\begin{equation}
\label{kappa_x}
\kappa_x=\frac{1}{2}\left[k_x\{\left\langle k_y^2\right\rangle-\left\langle 
k_z^2\right\rangle\}+\{\left\langle k_y^2\right\rangle-\left\langle 
k_z^2\right\rangle\}k_x\right]
\end{equation}
 and $\kappa_y$, $\kappa_z$ are obtained by cyclic permutations of $k_x$, $k_y$ 
and $k_z$. In the quantum wire, electrons can move only along $\hat x$ (the axis 
of the quantum wire). Hence setting $k_y=k_z=0$, the Dresselhaus Hamiltonian 
simplifies to
\begin{equation}
\label{dresselhaus2}
H_D(k)=\gamma\left(\left\langle k_y^2\right\rangle-\left\langle 
k_z^2\right\rangle\right) k_x\sigma_x  
\end{equation}
where $\left\langle k_y^2\right\rangle=\left(n\pi/W_y\right)^2$ and 
$\left\langle k_z^2\right\rangle=\left(m\pi/W_z\right)^2$. Here $m$ and $n$ are 
subband indices along $\hat z$ and $\hat y$, respectively. Also, $W_y$ and $W_z$ 
are wire dimensions along $\hat y$ and $\hat z$ respectively. 
Similarly, from equation (\ref{rashba1}) we can derive the  Rashba Hamiltonian 
to be
\begin{equation}
\label{rashba2}
H_R(k)=\eta\sigma_zk_x
\end{equation}

From equation (\ref{spin_evolution}) we can obtain the temporal evolution of the 
spin vector as \cite{bournel_1}, \cite{bournel_2}:
\begin{equation}
\label{temporal}
\frac{d\vec S}{dt}=\vec\Omega\times\vec S
\end{equation}
where the precession vector $\vec\Omega$ has two orthogonal components 
$\vec\Omega_R(k)$ and $\vec\Omega_D(k)$ due to Rashba and Dresselhaus 
interactions respectively:
\begin{subequations}
\begin{equation}
\label{omega_r}
\vec\Omega_R(k)=\frac{2a_{46}}{\hbar}E_yk_x\hat z
\end{equation}
\begin{equation}
\label{omega_d}
\vec\Omega_D(k)=\frac{2a_{42}}{\hbar}\left[\left(\frac{m 
\pi}{W_z}\right)^2-\left(\frac{n\pi}{W_y}\right)^2\right]k_x\hat x
\end{equation} 
\end{subequations}

Note that the precession vector $\vec\Omega$ lies in the $x-z$ plane (equations 
(\ref{omega_r}) and (\ref{omega_d})).
\begin{figure}
\centering
\includegraphics[width=2.5in]{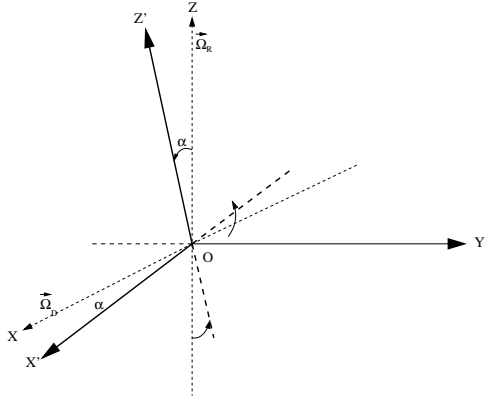}
\caption{Axis designations}
\label{spin1}
\end{figure}
Now we rotate the $x-z$ plane about the $y$ axis in a way (Figure \ref{spin1}) 
such that $\vec\Omega$ becomes coincident with the new $z$ axis. We name this 
new $z$ axis $z'$ and the new $x$ axis $x'$. This requires rotating the $x$ and 
$z$ axes through an angle $\alpha$ in the $x-z$ plane as shown in Figure 
\ref{spin1}. The angle $\alpha$ is given by 
\begin{equation}
\label{alpha}
\alpha=tan^{-1}\left(\frac{\Omega_D}{\Omega_R}\right)
\end{equation}

The spin precession equation (\ref{temporal}) in the $x'yz'$ coordinate system 
reads as follows:
\begin{subequations}
\begin{equation}
\label{evolution}
\frac{d\vec S}{dt}=\vec\Omega\times\vec S=det\left[\begin{matrix} \hat {x'}& 
\hat y & \hat {z'}\\
0&0&\Omega (t)\\
S_{x'}(t) & S_{y}(t) & S_{z'}(t) \end{matrix}\right] 
\end{equation} where 
\begin{equation}
\label{omega}
\Omega(t) =\sqrt{\Omega_D^2(t) +\Omega_R^2(t)} = \zeta k_x(t)
\end{equation}
and
\begin{equation}
\zeta = {{2}\over{\hbar}} \sqrt{ \left ( a_{46} E_y \right )^2 + a_{42}^2 \left 
[ \left ( {{m \pi}\over{W_z}} \right)^2 - \left ( {{n \pi}\over{W_y}} \right)^2 
\right ]^2}
\end{equation}
\end{subequations}
From equation (\ref{evolution}) we get
\begin{subequations}
\begin{equation}
\label{Sx'}
\frac{dS_{x'}}{dt}=-\Omega S_y
\end{equation}
\begin{equation}
\label{Sy}
\frac{dS_y}{dt}=\Omega S_{x'}
\end{equation}
\begin{equation}
\label{Sz'}
\frac{dS_{z'}}{dt}=0
\end{equation}
\end{subequations}
\begin{figure}
\centering
\includegraphics[width=2.5in]{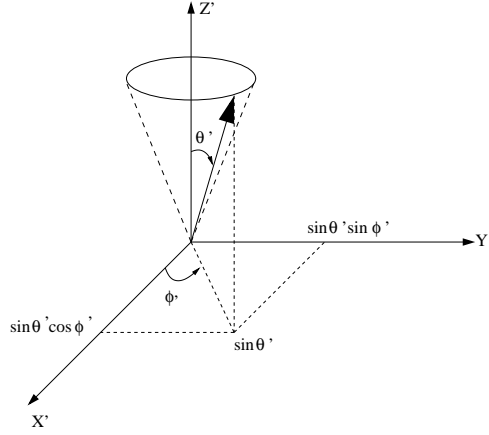}
\caption{Spin components in x'yz' co-ordinate system}
\label{spin2}
\end{figure}

In spherical co-ordinates (Figure \ref{spin2}), $S_{x'}=S\ sin\ \theta'cos\ 
\phi'$, $S_y=S\ sin\ \theta' sin\ \phi'$ and $S_{z'}= S\ cos\ \theta'$ where $S$ 
is the magnitude of the spin vector. Substitution of  these expressions in 
equations (\ref{Sx'}), (\ref{Sy}) and (\ref{Sz'}) yield
\begin{subequations}
\begin{equation}
\label{phi'}
\frac{d\phi'}{dt}=\Omega (t) = \zeta k_x(t)
\end{equation}
\begin{equation}
\label{theta'}
\frac{d\theta'}{dt}=0
\end{equation}
\end{subequations}
Solution of Equation (\ref{phi'}) yields
\begin{equation}
\begin{aligned}
\phi'(t) - \phi'(0) &= \zeta \int_0^t k_x(t') dt'\\ &= {{m^* \zeta}\over{\hbar}} 
\int_0^t v_x(t') dt'\\ &= {{m^* \zeta}\over{\hbar}} x \\ &\equiv \Phi' (x)
\label{angle}
\end{aligned}
\end{equation}
where we have assumed a parabolic electron energy dispersion so that the 
velocity is given by $v_x(t)$ = $\hbar k_x(t)/m^*$ ($m^*$ is the effective 
mass).
Before we proceed to derive the expressions for the spin components as a 
function of position $x$, we need to relate the primed quantities to their 
unprimed counterparts.  
\begin{equation}
\begin{aligned}
cos\ \theta' &= \frac{\vec S\cdot\hat z'}{|\vec S| |\hat z'|}\\ & =(S_x/S)\ z'_x 
+ (S_y/S)\ z'_y + (S_z/S)\ z'_z\\
&=(S_x/S)\ sin\ \alpha + (S_z/S)\ cos\ \alpha \\
&=cos\ \phi\ sin\ \theta\ sin\ \alpha + cos\ \theta\ cos\ \alpha
\label{trans1}
\end{aligned}
\end{equation}
where $z'_x$ = $sin\ \alpha$, $z'_y$ = 0 and $z'_z$ = $cos\ \alpha$.
From the above, we find
\begin{equation}
sin\ \theta' = \left[1-A\right]^{1/2}
\label{trans2}
\end{equation}
where
\begin{equation}
\begin{aligned}
A&=cos^2\theta'\\
&=\left(\frac{S_x}{S}\right)^2sin^2\alpha+\left(\frac{S_z}{S}\right)^2cos^2
\alpha+\frac{S_xS_z}{S^2}\ sin\ 2\alpha
\end{aligned}
\end{equation}

The components of $\vec S$ in the original system of coordinates $(x, y, z)$ are 
then easily obtained from the components in the primed system $(x', y, z')$.
\begin{equation}
\begin{aligned}
S_x&=S_{x'}\ cos\ \alpha+S_{z'}\ sin\ \alpha\\
&=S\ sin\ \theta' cos\ \phi' \ cos\ \alpha + S\ cos\ \theta' sin \ \alpha\\
S_y & = S_{y'}= S\ sin\ \theta' sin\ \phi' \\
S_z & = -S_{x'}\ sin\ \alpha+S_{z'}\ cos\ \alpha\\
& = -S\ sin\ \theta' cos\ \phi' \ sin\ \alpha + S\ cos\ \theta' cos \ \alpha\\
\end{aligned}
\label{spins}
\end{equation}
Using equations (\ref{trans1})--(\ref{spins}), we get
\begin{equation}
\begin{aligned}
S_x(x)&=S_0\ cos\ [\Phi'(x)+\phi'(0)]\ cos\ \alpha+ S_x(0)\ sin^2 \alpha\\ 
&+S_z(0)\ sin\ \alpha\ cos\ \alpha \\
S_y(x)&=S_0\ sin\ [\Phi'(x)+\phi'(0)]\\
S_z(x)&=-S_0\ cos\ [\Phi'(x)+\phi'(0)]\ sin\ \alpha\\&+ S_x(0)\ sin\ \alpha\ 
cos\ \alpha+ S_z(0)\ cos^2\ \alpha\\
\end{aligned}
\end{equation}
where 
\begin{equation}
\begin{aligned}
S_0^2&=S^2 - S_x(0)^2 sin^2 \alpha - S_z(0)^2 cos^2 \alpha 
\\&- S_x(0)\ S_z(0)\ sin\ 2 \alpha
\end{aligned}
\end{equation}

Let us now consider the situation where electrons are injected into the 
quantum wire with their spins polarized along the $+\hat x$ direction. In that 
case, $\phi'(0)=0$, $S$ = $S_x(0)$ and $S_y(0)=S_z(0)$ = 0. The above equation 
then simplifies to 
\begin{equation}
\begin{aligned}
S_x(x) & = S_x(0) \left [ cos^2 \alpha\ cos\ \Phi' + sin^2 \alpha \right ] \\
&=\frac{S_x(0)\left[\beta^2(m,n) +\left(a_{46}E_y\right)^2cos\ 
\gamma(m,n)x\right]}{\beta^2(m,n)+\left(a_{46}E_y\right)^2} \\
S_y(x) & = S_x(0)\  cos\ \alpha\ sin\ \Phi' \\
&=\frac{a_{46}\ E_y\ sin\ 
\gamma(m,n)x}{\sqrt{\beta^2(m,n)+\left(a_{46}E_y\right)^2}}\ S_x(0) \\
S_z(x) & = S_x(0)\ sin\ \alpha\ cos\ \alpha \left [ 1 - cos\ \Phi' \right ] \\
& =\frac{2\ a_{46}\ E_y\ \beta(m,n)\ 
S_x(0)}{\beta^2(m,n)+\left(a_{46}E_y\right)^2}\ sin^2 
\left[\frac{\gamma(m,n)x}{2}\right]
\end{aligned}
\label{relax}
\end{equation}
where
\begin{subequations}
\begin{equation}
\beta(m,n)=\frac{m^2\pi^2a_{42}}{W_z^2}- \frac{n^2\pi^2a_{42}}{W_y^2}
\end{equation}
\begin{equation}
\gamma(m,n)=\frac{2m^*}{\hbar^2}\sqrt{\beta^2(m,n)+\left(a_{46}E_y\right)^2}
\end{equation}
\end{subequations} 
It is straightforward to verify from equation (\ref{relax}) that 
\begin{equation}
\label{mod_S}
S_x(x)^2+S_y(x)^2+S_z(x)^2=1
\end{equation}
Thus, the magnitude of the spin vector is conserved only for every {\it 
individual electron}. However, when we have an ensemble of electrons, the 
magnitude of the {\it ensemble averaged} spin may decay with distance. This is 
the D'yakonov-Perel' relaxation. In the next section we investigate when this 
relaxation exists.
\section{Necessary conditions for D'yakonov-Perel' relaxation}
\subsection{Rashba interaction}
We can see immediately from equation (\ref{relax}) that if there is no Rashba 
interaction ($a_{46}$ = 0 or, $E_y$ = 0), then {\it at all positions $x$},
\begin{equation}
\begin{aligned}
\left\langle S_x(x)\right\rangle & =\left\langle S_x(0)\right\rangle \\
\left\langle S_y(x)\right\rangle & = \left\langle S_z(x)\right\rangle = 0 
\end{aligned}
\end{equation}
Therefore,
\begin{equation}
\label{new}
\begin{aligned}
\left|\left\langle\vec S(x)\right\rangle\right| &=\sqrt{{\left\langle 
S_x(x)\right\rangle^2 + \left\langle S_y(x)\right\rangle^2 + \left\langle 
S_z(x)\right\rangle^2}}\\
& = \left\langle S_x(0)\right\rangle \\ 
&= \left|\left\langle{\vec S}(0)\right\rangle\right| \\
& = a~constant~independent~of~position~x
\end{aligned}
\end{equation}
Here the angular brackets $\langle~\rangle$ denote ensemble average over 
electrons and $\langle\vec S(x)\rangle$ is the ensemble averaged spin vector at 
position $x$.

Equation (\ref{new}) indicates that as along as the carriers are injected with 
their spins aligned 
along the axis of the wire, there is {\it no D'yakonov-Perel'} relaxation, since 
the ensemble average spin $|\langle{\vec S}\rangle|$ does not decay at all. 
Therefore, 
{\it{Rashba interaction is required}} for the ensemble averaged spin to relax.

\subsection{Dresselhaus interaction}
If there is no Dresselhaus interaction ($a_{42}$ = 0), then
\begin{equation}
\begin{aligned}
\left\langle S_x(x)\right\rangle & =\left\langle S_x(0) cos \left [ \left ( {{2 
m^* a_{46}E_y}\over {\hbar^2}} \right ) x \right ]\right\rangle \\
\left\langle S_y(x)\right\rangle & = \left\langle S_x(0) sin \left [ \left ( {{2 
m^* a_{46}E_y}\over {\hbar^2}} \right ) x \right ]\right\rangle \\
\left\langle S_z(x)\right\rangle & = 0 \nonumber \\
\end{aligned}
\end{equation}
Therefore,
\begin{equation}
\begin{aligned}
\left|\left\langle{\vec S}(x)\right\rangle\right| & =  \sqrt{{\left\langle 
S_x(x)\right\rangle^2 + \left\langle S_y(x)\right\rangle^2 + \left\langle 
S_z(x)\right\rangle^2}} \\
&= \left|\left\langle{\vec 
S}(0)\right\rangle\right| \nonumber \\
& = a~constant~independent~of~position~x
\end{aligned}
\end{equation}
Again, we see that the ensemble averaged spin $|\langle{\vec S}\rangle|$ does 
not decay. In
this case, the spin oscillates between the $x$- and $y$-polarization (the 
$z$-polarization remains $0$), but the ``amplitude'' of this oscillation does 
not decay. Therefore, {\it{there can be no D'yakonov-Perel' relaxation without 
Dresselhaus interaction.}}

\subsection{Multi-channeled transport}

If both Rashba and Dresselhaus interactions are present, but transport is single 
channeled, i.e. $m$ = $m_0$ and $n$ = $n_0$, then {\it every} electron 
is in the same subband ($m_0$, $n_0$). In that case, 
\begin{equation}
\begin{aligned}
\left\langle S_x(x)\right\rangle^2 & = \left[\frac{S_x(0)\left(\beta^2_0 
+(a_{46}\ E_y)^2cos\ \gamma_0x\right) }{\beta^2_0 
+\left(a_{46}\ E_y\right)^2}
\right]^2 \\
\left\langle S_y(x)\right\rangle^2 & = \left [ \frac{a_{46}\ E_y\ S_x(0)\ sin\ 
\gamma_0x}{\sqrt{\beta^2_0+\left(a_{46}\ E_y\right)^2}} \right ]^2\\
\left\langle S_z(x)\right\rangle^2 & = \left [\frac{2\ a_{46}\ \beta_0\ E_y\ 
S_x(0)\ sin^2 \left(\gamma_0x/2\right)}{\beta^2_0 +\left(a_{46}\ E_y\right)^2} 
\right]^2
\end{aligned}
\end{equation}
where $\beta_0=\beta(m_0,n_0)$, and $\gamma_0=\gamma(m_0,n_0)$.

Once again, it is easy to verify that 
\begin{equation}
\begin{aligned}
|\langle{\vec S}(x)\rangle|& = \sqrt{{\left\langle S_x(x)\right\rangle^2 + 
\left\langle S_y(x)\right\rangle^2 + \left\langle S_z(x)\right\rangle^2}}\\ 
&= |\langle{\vec S}(0)\rangle| \nonumber \\
& = a~constant~independent~of~position~x
\end{aligned}
\end{equation}
Consequently, there is no D'yakonov-Perel' relaxation if transport is {\it 
single channeled}. This is true regardless of whether the electrons are injected 
into the lowest
subband, or any other subband, as long as there is no inter-subband transition.

\section{What is necessary for D'yakonov-Perel' relaxation?}

If transport is multi-channeled,  then different electrons 
at position $x$ are in different subbands. In that case, the indices $n$ 
and $m$ are different for different electrons, so that  ensemble
averaging results in
\begin{equation}
\begin{aligned}
\left|\left\langle{\vec S}(x)\right\rangle\right| &=  \sqrt{{\left\langle 
S_x(x)\right\rangle^2 + \left\langle S_y(x)\right\rangle^2 + \left\langle 
S_z(x)\right\rangle^2}}\\
 &\neq \left|\left\langle{\vec S}(0)\right\rangle\right|
\end{aligned}
\end{equation} 

Therefore, multi-channeled transport, in the presence of {\it both}
Rashba and Dresselhaus interaction leads to D'yakonov-Perel' relaxation.
It is important to note that ``scattering'', or inter-subband transitions 
are {\it{not required}} for the D'yakonov-Perel relaxation. Even if every 
electron
remains in the subband in which it was originally injected, there will be a 
D'yakonov-Perel' relaxation as a consequence of {\it ensemble averaging}
over the electrons. Of course, if there is scattering and inter-subband 
transitions, then the subband indices ($m$, $n$) for every electron 
becomes a function of position $x$, in which case the effect of ensemble 
averaging is exacerbated and the relaxation will be more rapid. Thus, we have 
established that three conditions are needed for D'yakonov-Perel' relaxation:
(i) Rashba interaction, (ii) Dresselhaus interaction, and (iii) multi-channeled 
transport. \par
\begin{figure}
\centering
\includegraphics[width=2.5in]{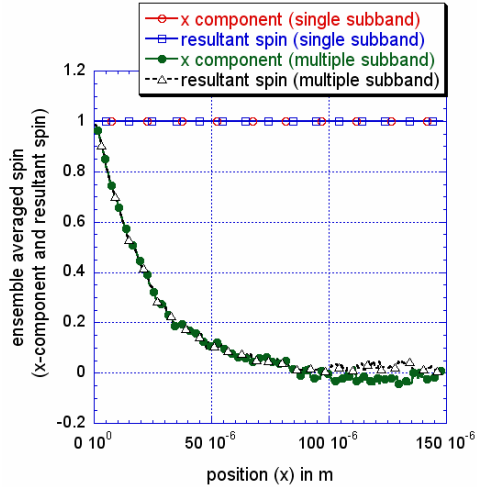}
\caption{Spin relaxation in a $GaAs$ quantum wire of rectangular cross section 
$30$nm$\times4$nm. The driving electric field is 2kV/cm and lattice temperature 
is $30$K. These results are obtained from Monte Carlo simulation described in 
\cite{pramanik_prb}, \cite{pramanik_apl}. Spin does not relax for single 
channeled (single subband) transport, but does relax for multichanneled 
transport.}
\label{comparison}
\end{figure}

 In Figure \ref{comparison}, we show $\left|\langle\vec 
S(x)\rangle\right|$ as a function of $x$ for two cases: single channeled 
transport and multi-channeled transport. It is evident that the spin does not 
decay for single channeled transport but does decay for multi-channeled 
transport.

\section{Conclusion}

In this paper, we have established the origin of the D'yakonov-Perel' spin 
relaxation in a quantum wire.
 This relaxation is harmful for most spintronic 
devices (one example is the SPINFET \cite{datta}), because it leads to spin 
randomization. Since  optimum materials for SPINFET-type devices (e.g InAs) 
usually possess strong Rashba and 
also some Dresselhaus spin orbit interactions, the only effective 
way to eliminate the D'yakonov-Perel' relaxation is to ensure and enforce 
single channeled transport. There has been recently some proposals that advocate 
using multi-channeled devices for SPINFET's, along with the claim that they 
provide better spin control via the use of multiple gates \cite{loss}. While we 
do not believe that spin control is improved by using multiple gates since 
synchronizing these gates is an additional engineering burden that can only 
degrade device operation and gate control, it is even more important to 
understand that multi-channeled devices
have serious drawbacks. The original proposal for the SPINFET  pointed out that 
multi-channeled transport is harmful because it dilutes the spin interference 
effect which is the basis of the SPINFET device \cite{datta}. 
Here, we have pointed out an additional motivation to avoid multi-channeled 
devices: they will suffer from D'yakonov-Perel' relaxation, while the single 
channeled device will not.

\end{document}